\documentclass[10pt]{article}
\usepackage{amssymb}
\usepackage{amsmath}
\usepackage{wasysym}
\usepackage{graphicx}

\begin{document}

\title{Backward non-unitary quantum evolution}

\author{C. Dedes\thanks{{c\_dedes@yahoo.com}}\\
Bradford College  \\
Great Horton Road, Bradford \\
West Yorkshire, BD7 1AY \\
United Kingdom \\}

\maketitle

\begin{abstract}
A non-linear backward equation with diffusive terms is postulated for the probability density that depends on the Bohmian quantum potential. An associated nonlinear Schr\"{o}dinger equation is also introduced and extension of the analysis to several particle compounds is sketched along with the implications following from this formalism regarding the non-conservation of probability in the non-equilibrium regime. Some further conclusions are educed with respect to the generalized optical theorem.
\end{abstract}
 
\section{Introduction}

In one of the early papers on wave mechanics, Darwin \cite{Darwin,deBroglie} examined an interesting example of a two-particle quantum interference involving a set up of a coherent pair of particles. A few years later interparticle quantum correlations between two bodies became notable with the work of Einstein, Podolsky and Rosen\cite{EPR,Bohr} and since then have been at the heart of quantum mechanics. The issue of non-separability between interacting particles had been particularly emphasized by Einstein in his own version of the EPR thought experiment \cite{Howard}. Schr\"{o}dinger in a similar vein claimed quite succinctly that according to quantum theory maximum knowledge of a composite system does not necessarily mean fullest knowledge of the individual particles \cite{Schrodinger,Schrodinger2}. This type of reasoning that emphasizes the influence of the particular experimental conditions for the available type of possible predictions has been extensively used by Bohr in his reply to Einstein-Podolsky-Rosen \cite{Bohr} and is best illustrated in the ingenious thought experiments devised earlier by him and Rosenfeld for the measurability of fields \cite{BohrRosenfeld}. More recently Mermin in particular emphasized the physical reality of these subsystem correlations in comparison to the \textit{correlata} underlying them. From a rather different perspective, the quantum potential which has a pre-eminent role in Bohmian dynamics, is associated with the interrelatedness in the quantum world and suggests a quality of irreducible wholeness in the domain of quantum phenomena \cite{BohmHiley,BohmHiley2}. Here we will modify the continuity equation for the probability density by including a term proportional to the quantum potential and examine some of its consequences.

Another issue that needs to be addressed is the time asymmetry observed in the physical world in contradistinction to the time symmetrical form of the fundamental equations of evolution \cite{Sklar}. The resolution of the contradiction in classical mechanics is usually related to an asymmetry in initial conditions. In quantum mechanics in particular, Schr\"{o}dinger equation becomes time reversible if we make the identification of the probability amplitude with its time symmetrical complex conjugate one. It is generally appreciated that the quantal irreversibility is the result of initial conditions or the outcome of the interaction of a few particle system with a reservoir but recently it has been shown experimentally \cite{Paternostro} that even a few particles may exhibit signs of irreversibility. The collapse of the wavefunction on the other hand is generally accepted to impose a genuinely time asymmetric element in the quantum formalism. 

In this work we follow a different approach, by demonstrating that the quantum potential term mentioned above may be employed in a way that implies that the evolution equation for the probability density is genuinely time-asymmetric. It deserves notice that the equation that will be proposed depends on a final value condition  so it turns out to be backward \cite{Ockendon} and resembles on that respect the Kolmogorov backward equation \cite{Gardiner}, the adjoint transport equation \cite{Nuclear} or the Black-Scholes one known from mathematical finance \cite{Wilmott}. Following this we extend our study for the multiparticle and present some conclusions related to the free-particle problem and the generalized optical theorem and finally we recapitulate our findings in the final section.

\section{Theory}

\textit{Generalized continuity equation and its associated non-linear wave equation.}
In a recent paper \cite{Renner} it has been argued that quantum theory may not be extended straightforwardly to the macroscopic domain. This could lead us to seek generalizations of the present quantum formalism. 

From a more fundamental perspective the tenseless world view implied by the time symmetry of the quantum mechanical formalism, which is usually regarded as a virtue of the theory, may be in fact a severe limitation and a byproduct of its highly abstract nature. The same could be argued about unitary evolution which is unable to account for observed phenomena in statistical and condensed matter physics \cite{Drossel}. We tend to embrace a quite different opinion according to which a lawlike temporal asymmetry would be instead a \textit{desideratum} in a possible generalization of standard quantum theory. It been rightly claimed that the existence of a classical world is essential and even a precondition for the consistency of the orthodox quantum interpretation, in the same manner we argue that temporal passage is a prerequisite for any physical theory that attempts to capture the essentials features of the natural world and more general aspects of the human experience \cite{Gisintime}. The so-called spatialization of time has some times obscured the fact that even to start making sense of an external physical world we must presuppose dynamic change and consequently a conceptual framework that admits a sharp distinction between past and future \cite{Feser}.   

Since standard non-linear versions of the Schrodinger equation \cite{Birula,Weinberg2} seem to face certain difficulties \cite{Gisin} we try instead to generalize the Bohm-de Broglie formulation. Employing the position representation, we consider initially a neutral, spinless  particle of mass $m$, assuming that the time varying potential is $V(t)$. The fundamental quantum mechanical equation of motion for the single particle probability amplitude reads as

\begin{equation}
i\hbar\frac{\partial \Psi}{\partial t}=\left(-\frac{\hbar ^{2}}{2m}\nabla ^{2}+V\right)\Psi .
\end{equation}

\noindent
The familiar polar substitution $\Psi=Re^{iS/\hbar}$ gives the  Hamilton-Jacobi-Madelung equation when we equate the real parts

\begin{equation}
    \frac{\partial S}{\partial t}=\frac{\hbar ^{2}}{2m}\left[\frac{(\nabla^{2}R)}{R}-\frac{1}{\hbar^{2}}(\nabla S)^{2}\right]-V.
\end{equation}

\noindent
We may also define the quantum potential as 

\begin{equation}
    Q=-\frac{\hbar ^{2}}{2m}\frac{\nabla^{2}R}{R},
\end{equation}

\noindent
which is important in Bohm's version of the de Broglie-Bohm mechanics, although it is possible to formulate the theory without any reference to it. Equating the imaginary parts we obtain

\begin{equation}
  \frac{\partial R}{\partial t}+\frac{1}{2m}\left(2\nabla R\cdot\nabla S+R\nabla^{2}S\right)=0.
\end{equation}

\noindent
Born's rule which at least in text-book quantum mechanics is considered a separate axiom of quantum physics provides a conceptual and calculational bridge between theory and experimental results. Taking into account then the identification $\rho=u^{2}$ for the probability density and multiplying both sides with $2u$ yields the familiar continuity equation

\begin{equation}
    \frac{\partial \rho}{\partial t}+\nabla\cdot\left(\rho\frac{\nabla S}{m}\right)=0.
\end{equation}

\noindent
From (2) and (4) we may determine in principle the action S and the amplitude R. We seek to modify the previous expression (4) including a diffusion term and a linear term proportional to the quantum potential as follows
 
\begin{equation}
  \frac{\partial u}{\partial t}+\frac{1}{2m}\left(2\nabla u\cdot\nabla S+u\nabla^{2}S\right)=\left(-\frac{\hbar}{2m}\right) \left(\nabla^{2}u-\frac{\nabla ^{2}R}{R}u\right).  
\end{equation}

\noindent
This formula may be considered as a non-equilibrium extension of (4) in the sense that as $u\rightarrow R$ the former reduces to the familiar equation that holds at quantum equilibrium. The given expression remains time-asymmetrical which immediately implies non-unitary evolution (the inverse proposition is not true nevertheless). The negative coefficient suggests a backward evolution for $u(t)$. We conclude then that not only the evolution equation is not time-invariant but that also there is time asymmetry in the imposition of terminal instead of initial conditions. An element then of finality is introduced in (6) since it relies on terminal values evolving backwards in time. The homogeneity condition is satisfied since for an arbitrary solution $u$ and $\kappa$ a real number it follows immediately that $\kappa u$ is also a solution. Hence the rescaling transformation $u'\rightarrow\kappa u$ constitutes a similarity transformation that leaves the equation invariant. An imaginary quantum potential would give rise to dissipative terms that do not preserve unitarity in the general case. At the same time we must ensure that the expression reduces to the standard form at equilibrium. This suggests an extension of (1)  written as

\begin{equation}
i\hbar\frac{\partial \Phi}{\partial t} = \left(-\frac{\hbar ^{2}}{2m}\nabla ^{2}+V\right)\Phi -i\frac{\hbar^{2}}{2m}\left(\frac{\nabla^{2}\sqrt{|\Phi|^{2}}}{\sqrt{|\Phi|^{2}}}-\frac{\nabla^{2}\sqrt{|\Psi|^{2}}}{\sqrt{|\Psi|^{2}}}\right)\Phi.  
\end{equation}

\noindent
where $\Phi=ue^{iS/\hbar}$ and the wavefunction $\Psi=Re^{iS/\hbar}$ is governed by (1) with appropriate final conditions. These kind of polar substitution lead immediately to (6) and a Hamilton-Jacobi equation identical to (2) with $R$ replaced by $u$. The existence of a distinct wave function with the same phase as $\Psi$ but different real amplitude is reminiscent of the double solution theory put forward by de Broglie \cite{Durt}, but it must be noted that in (7) $\Phi$ propagates backward in time in the configuration space in the same manner as $\Psi$, so it is not a singularity in the real three dimensional space as de Broglie envisaged. Dynamical terms of the form given in (7) have also been used in the past in the real but not in the imaginary part of the Hamiltonian  \cite{Smolin}. It must be underscored that when the terminal condition is fulfilled, which means $u(t\rightarrow \infty)=R(t\rightarrow \infty)$ so $\Phi(t\rightarrow \infty)=\Psi(t\rightarrow \infty)$ the superposition principle holds and we recover (1) from (7) even though the coefficient $\hbar/m$ is obviously different from zero. Hence the magnitude of the non-linearity coefficient may be non-zero nevertheless there is an appropriate limit that we recover (1). On the other hand a nonlinear Schr\"{o}dinger as in \cite{Birula} reduces to the linear one only when the  nonlinearity constant becomes zero. Finally, (7) is time irreversible even when $\Phi\rightarrow \Phi^{*}$ and respects the homogeneity condition \cite{Weinberg2} which says that if $\Phi$ is a solution of (7) so is $\Phi\rightarrow Z \Phi$, where $Z$ a constant complex number. Based on (7) we can present extensions of standard quantum formulas. If we write for example the expectation value of an operator $\hat{A}$ as $\langle \hat{A} \rangle=\int \Phi^{*} \hat{A} \Phi  d^{3}$ its time development is given by

\begin{equation}
  \frac{d\langle \hat{A} \rangle}{dt}=\frac{1}{ih}\langle [\hat{A},\hat{H}] \rangle+\left \langle \frac{\partial \hat{A}}{\partial t} \right \rangle-\frac{\hbar}{m}\mathbf{Re}
  \int \Phi^{*} \left(\frac{\nabla^{2}\sqrt{|\Phi|^{2}}}{\sqrt{|\Phi|^{2}}}-\frac{\nabla^{2}\sqrt{|\Psi|^{2}}}{\sqrt{|\Psi|^{2}}}\right)\hat{A} \Phi  d^{3}x,
\end{equation}

\noindent
where $\hat{H}$ the Hamiltonian and of course the first two terms make up the familiar expression in the Schr\"{o}dinger picture whereas the last term reflects dissipative effects. According to (6) and taking into account that $\rho=|\Phi|^{2}$, the corresponding expression for the probability density is

\begin{equation}
 \frac{\partial \rho}{\partial t}+\nabla
\cdot\left(\rho\frac{\nabla S}{m}\right)= \left(-\frac{\hbar}{m}\right) \left[\frac{1}{2}\nabla^{2}\rho-\frac{(\nabla \rho)^{2}}{4\rho}-\frac{\nabla ^{2}R}{R}\rho\right]. 
\end{equation}

\noindent
The left part is familiar from the continuity equation but the right non zero part implies non-unitary evolution. The existence of these terms suggests a possible interplay between spatial-temporal coherence and diffusion phenomena. Diffusion terms in the continuity equation have been also been added in \cite{Goldin,DG} (see also \cite{Schuch,Kostin}). In addition equation (9) remains invariant when $R\rightarrow\kappa R$ and $\rho\rightarrow \lambda\rho$. It must be noted though that the diffusion constant in our proposal is negative, namely $-\hbar/m$ as we mentioned. Without the nonlinear and the quantum potential term, (9) is a non-equilibrium backward Fokker-Planck type equation that conserves probability locally. The inclusion of the quantum potential and nonlinear terms destroy this property and is responsible for its non-conservation. It is evident that superluminal communication is precluded by the very fact that the relevant equation of motion for the probability density depends on the imposition of a final condition. This preferential status of terminal conditions instead of initial means that we could use (6),(7) or (9) to make retrodictions but not predictions. It may be commented that this special importance of final conditions can be clearly illustrated as with the help of the well-known thought delayed choice experiment \cite{Wheeler} and even the Bohrian emission of photons during atomic transitions in the old quantum theory era. 

Propagation in the negative time direction is not something novel, a prominent example in classical physics is the advanced electromagnetic potential, a quantity itself not directly measurable, which is a solution of time symmetric equations. In the present case we deal instead with nonlocal evolution equations with explicit time asymmetry. The question that naturally arises is if we can exploit at least in principle these features in order to send superluminal signals. In the presented framework this can only be attempted in the backward time direction. In order to do this we need to make use of some causal link between future and past states, always bearing in mind that according to (6) or (9) any possible influence is strictly of statistical nature. In more concrete terms this could be achieved for example by the placement (or absence) of a detector or another device and question its possible influence for earlier times. This situation parallels the aforementioned delayed choice experiment. Wheeler did not proclaim a violation of causality or influence to past dynamics but concluded instead that there is no definite past unless it is observed in the present. Even though we advance a different interpretation, this assertion already introduces an element of end-directedness or physical intentionality from past to present states and suggests a kind of teleological causality. We refer the reader to a contemporary discussion on teleological causation, without excessive Aristotelian commitments, by Hawthorne and Nolan \cite{Hawthorne} (especially the particularly illuminating section on teleology and backward causation and the comparison between teleology and non-locality). It is important to distinguish between backward efficient causation and (forward) teleological or final causation because only the former could be useful for signalling. In order to do this and following these authors we might involve the notion of an abstract distance from a privileged end state. In our treatment this privileged end would be $R$ in (6) and $u=R$ is the equilibrium limit. We notice that the equation of motion for $u$ depends on the privileged end $R$ which is perfectly consistent in a final cause context. In contradistinction in the backward efficient cause case one should proceed in what these authors call 'stepwise fashion' and a connecting causal chain. What is of crucial importance is that the final cause case remains valid even if the system does not actually reach this privileged end. We can then rule out the backward efficient causation and favour the teleological explanation which at the same time ensures that no signalling is possible.

At the equilibrium limit equations (6) and (9) reduce to the more familiar Bohmian equations so they may been seen as generalizations of the latter in the non-equilibrium domain. The presence of the diffusion terms would tend to make $u$ smoother but the quantum potential driving term and especially the highly nonlinear midterm could have the opposite effect. On physical grounds we would expect irregularities in $\rho (x,t)$ to become less smooth with backward time evolution which may cause problems for the behaviour of the solutions like arbitrary rapid growth in finite time \cite{Ockendon}. The presence then of the nonlinear terms is of special importance, since if they tend to create irregularities as we propagate backwards in time and dominate over the decay term that will decide in favour of the ill-posedeness of the problem and would demand sophisticated regularization techniques. Obviously, we could also include advanced vector potentials and even noise terms in (9) \cite{RMP}. The latter ones require special care though because of certain mathematical subtleties involved \cite{Hairer}. 

 It must be emphasized though that there are no non-intersecting particle trajectories apart from the case of equilibrium when (7) becomes the familiar density current equation and the coefficient sets time scale equal to $\hbar/m$. We now seek a guidance equation modifying the de Broglie-Bohm one. Motivated by (9) we write a formula for the velocity $\mathbf{v}$ of a single particle that would accompany (7) as

\begin{equation}
  \mathbf{v}=\frac{\hbar}{m}\mathbf{Im}\frac{\Phi^{*}\nabla\Phi}{|\Phi|^{2}}-\frac{\hbar}{2m}\left[\nabla ln |\Phi|^{2}-\frac{1}{|\Phi|^{2}}\int d^{3}x\left(\frac{(\nabla |\Phi|^{2})^{2}}{2|\Phi|^{2}}+\frac{2|\Phi|^{2}\nabla ^{2}\sqrt{|\Psi|^{2}}}{\sqrt{|\Psi|^{2}}}\right)\right].
\end{equation}

\noindent
The first term in the above formula is the velocity term $\frac{\nabla S}{m}$, the second one proportional to $\nabla ln |\Phi|^{2}$ is an osmotic type contribution to velocity according to \cite{BohmHiley,Groessing,Durt2} (note the negative sign of the diffusion coefficient) and the third integral term expresses the non-local character of this generalized guidance equation and clearly violates the uniqueness property of trajectories. Since the last two term cancel each other when $\Phi=\Psi$ we recover the standard guidance equation.

 \textit{Many particle case.} It is straightforward to  extend our previous results for the case of a compound of several interacting particles of equal masses $m$

\begin{equation}
 \frac{\partial \rho}{\partial t}+\sum _{i=1} ^{N}\nabla _{i}\cdot\left(\rho\frac{\nabla _{i}S}{m}\right)=\left(-\frac{\hbar}{m}\right)\sum _{i=1} ^{N}  \left[\frac{1}{2}\nabla^{2}_{i}\rho-\frac{(\nabla _{i} \rho)^{2}}{4\rho}-\frac{\nabla _{i} ^{2}R}{R}\rho\right]. 
\end{equation}

\noindent
The presence of the many particle quantum potential in the continuity equation signifies an important feature, namely the dependence of the multiparticle density on a quantity that exhibits non-local characteristics (but since (11) is backward in time there is no possibility of non-local signalling).

The interconnectedness of the system constituents is expressed then through the form of this non-local quantum potential. The constituent particles then are reidentified and incorporated into a substantial whole unified by a single unifying principle and cannot be considered as distinct aggregated components. This means that the component parts even though are retrievable in principle, they are not intrinsically unaffected by their union \cite{Scaltsas,Oderberg,Feser}. Similar considerations in a broader sense have also been expressed various authors. Whitehead for example has spoken about the concept of a dynamic and contextual organic whole \cite{Whitehead}, as of course Bohm and Hiley \cite{Bohm,Hiley} already mentioned. Having established the basic elements of the formalism we proceed by examining some further developments and examples.

\section{Examples}
\textit{Free quantum particle.} We consider a free non-relativistic particle of mass $m$ in one spatial direction. As it is well known the solution of the Schr\"{o}dinger equation is a plane wave and the probability density is constant with no spatial or temporal dependence. We can confirm by direct substitution that a plane wave solution (or a superposition integral of plane waves) does not satisfy (7). So we already conclude that a possible solution will dictate a less irregular pattern for the probability density as it propagates in the backward direction of time. As it was stated as a theoretical postulate in the previous section the phase of the function $\Phi$ will identical to the one that obeys the linear equation. Since $\Psi=Ae^{i(kx-\omega t)}$ then it follows that $\Phi=u(x,t)e^{i(kx-\omega t)}$. The two solutions have identical final values but as we will see they diverge for earlier times and $u(x,t)$ acquires spatial and time dependence. Inserting the expressions for $\Phi,\Psi$ in polar form or using (6) gives

\begin{equation}
 \frac{\partial u(x,t)}{\partial t}=
    -\frac{\hbar}{2m}\left(\frac{\partial ^{2} u(x,t)}{\partial x^{2}}+2k\frac{\partial u(x,t)}{\partial x} \right).
\end{equation}

\noindent
 We make the substitution $\tau=T-t$ and $T$ is the final time and noticing that the coefficients in the above equation are constant it follows that through a transformation of the type \cite{Polyanin}

\begin{equation}
    u(x,t)=e^{-kx-\left(\frac{\hbar k}{2m}\right)^{2}\tau}w(x,\tau),
\end{equation}

\noindent
we can reduce it to a readily solved heat equation 
    
    \begin{equation}
      \frac{\partial w(x,\tau)}{\partial \tau}=
    \frac{\hbar}{2m}\frac{\partial ^{2} u(x,\tau)}{\partial x^{2}}.  
    \end{equation}
    
\noindent
The familiar form of the heat equation along with final conditions means that our problem is well-posed but as explained in the previous section we should be aware that we must not use (14) if we wish to determine $u(x,\tau ')$ from $u(x,\tau)$ if $\tau'< \tau$ because this is an ill-posed problem. It follows from (12) or (14) that in the limiting case where the diffusion constant $\hbar/m$ tends to zero, $u(x,\tau)$ becomes constant which is the familiar result. The fundamental solution of (12) is a modified heat kernel of the form \cite{Polyanin}

\begin{equation}
   \mathcal{K}(x,\tau)=\frac{1}{\sqrt{\frac{2\pi\hbar}{m}\tau}}e^{-\frac{\left(x+\frac{\hbar k}{m}\tau\right)^{2}}{\frac{2\hbar}{m}\tau}}.
\end{equation}

\noindent
This solution becomes more regular as it propagates in the backward time direction nevertheless it is always more irregular in comparison to the constant amplitude of a plane wave. Based on (15) we can obtain solutions depending on appropriate boundary conditions and for spatially periodic problems the fundamental solution can be expressed as theta Jacobi function. It us well known that the heat kernel in the familiar framework of forward time evolution implies infinite propagation velocity but interestingly enough in the present context this does not pose any problems related to instant signalling. At this point it must be underlined that the point of view adopted in this work is that the backward propagation entails a certain directedness towards future events and allows retrodiction but this does not imply influence to the past and should not be equated to backward causation \cite{Horwich}. We conclude then that even the free particle problem is non-trivial and involves features of thermalization of the standard plane wave solution. The slightly more complicated case of a time-independent Gaussian wavefunction can be handled along the same lines and leads to qualitatively similar results, namely reduction to a backward heat equation and hence thermalization of the wavefunction. More general problems where the quantum potential has both spatial and time-dependence deserve are more involved and deserve further attention in order to justify or not their well-posedness.    

\textit{Quantum scattering and the generalized optical theorem.} Up to this point we have only considered time dependent evolution equations. It is possible nevertheless to make use of a time-independent version of (7) invoking its imaginary potential. The time-independent variant of this nonlinear wave equation for a particle with energy $E$, mass $m$ and wave-vector $\mathbf{k}$ is written as

\begin{equation}
   \left[\left (\nabla^{2}+k^{2}-U_{r}\right)-i\left(\frac{\nabla^{2}\sqrt{|\Phi(\mathbf{r})|^{2}}}{\sqrt{|\Phi(\mathbf{r})|^{2}}}-\frac{\nabla^{2}\sqrt{|\Psi(\mathbf{r})|^{2}}}{\sqrt{|\Psi(\mathbf{r})|^{2}}}\right)\right]\Phi(\mathbf{r})=0
\end{equation}

\noindent
where $k$ the wavenumber, $U(\mathbf{r})=\frac{2m}{\hbar ^{2}}V(\mathbf{r})$ and $V(\mathbf{r})$ a central potential. When $\Phi(\mathbf{r})=\Psi(\mathbf{r})$ the formal solution of (16) can be written as

\begin{equation}
    \Psi (\mathbf{r})=\Psi_{0}(\mathbf{r})+\int G_{\mathbf{k}}(\mathbf{r},\mathbf{r}') U(\mathbf{r}') \Psi (\mathbf{r}')d^{3}x
\end{equation}

\noindent
where $G_{\mathbf{k}}(\mathbf{r},\mathbf{r}')$ the corresponding Green function and $\Psi_{0}(\mathbf{r})$ a solution of the homogeneous equation $(\nabla^{2}+k^{2})\Psi_{0}(\mathbf{r})=0$ \cite{Wu}. Due to the nonlinear character of (16) it is not possible to write a relation similar to (17) for $\Phi(\mathbf{r})$. We can proceed though by assuming that at large distances from the potential region the solution $\Phi(\mathbf{r})$ has the asymptotic form 

\begin{equation}
    \Phi_{\mathbf{k}} \rightarrow  e^{i\mathbf{k}\cdot\mathbf{r}}+\frac{e^{ikr}}{r}f(\mathbf{k},\mathbf{k}').
\end{equation}

\noindent
Following the analysis by Glauber \cite{Glauber,Wu} for two scattering vectors $\mathbf{k},\mathbf{k}'$ (16) produces the identities

\begin{equation}
   \Phi_{-\mathbf{k}'} \nabla ^{2}\Phi_{\mathbf{k}}-\Phi_{\mathbf{k}'} \nabla ^{2}\Phi_{-\mathbf{k}}=0
\end{equation}

\noindent
and

\begin{equation}
   \nabla \cdot \left[\Phi^{*}_{\mathbf{k}'}\nabla\Phi_{\mathbf{k}}-\Phi_{\mathbf{k}}\nabla\Phi^{*}_{\mathbf{k}'}\right]+2i\left(\frac{\nabla^{2}\sqrt{|\Phi_{\mathbf{k}}|^{2}}}{\sqrt{|\Phi_{\mathbf{k}}|^{2}}}-\frac{\nabla^{2}\sqrt{|\Psi_{\mathbf{k}}|^{2}}}{\sqrt{|\Psi_{\mathbf{k}}|^{2}}}\right)\Phi _{\mathbf{k}'}^{*}\Phi_{\mathbf{k}}=0.
\end{equation}

\noindent
Integrating (19) and using Green's theorem over the volume of a very large sphere gives $f(\mathbf{k}',\mathbf{k})=f(-\mathbf{k}',-\mathbf{k})$ which expresses the reversibility of the scattering processes/amplitudes for two directions \cite{Glauber}. Furthermore, if the scattering potential has inversion symmetry then the scattering amplitude is symmetric  $f(\mathbf{k}',\mathbf{k})=f(\mathbf{k},\mathbf{k}')$ which is clearly related to the principle of detailed balance \cite{Glauber}. This conclusion nevertheless does not contradict our previous discussion which relied on a time-dependent evolution equation. Following the same procedure for (20) we find

\begin{eqnarray}
   \frac{1}{2i}[f(\mathbf{k}',\mathbf{k})-f^{*}(\mathbf{k},\mathbf{k}')]=\frac{k}{4\pi}\int f^{*}(\mathbf{k}'',\mathbf{k}) f(\mathbf{k}'',\mathbf{k})d\Omega_{\mathbf{k}''} \nonumber \\ -\frac{1}{4\pi}\int\left(\frac{\nabla^{2}\sqrt{|\Phi_{\mathbf{k}}|^{2}}}{\sqrt{|\Phi_{\mathbf{k}}|^{2}}}-\frac{\nabla^{2}\sqrt{|\Psi_{\mathbf{k}}|^{2}}}{\sqrt{|\Psi_{\mathbf{k}}|^{2}}}\right)\Phi_{\mathbf{k}}\Phi^{*}_{\mathbf{k}'}d^{3}x.
\end{eqnarray}

\noindent
Setting $\mathbf{k}=\mathbf{k}'$ gives

\begin{equation}
    \mathbf{Im} f(\mathbf{k},\mathbf{k}')=\frac{k}{4\pi}\sigma_{sc}-\frac{1}{4\pi}\int \left(\frac{\nabla^{2}\sqrt{|\Phi_{\mathbf{k}}|^{2}}}{\sqrt{|\Phi_{\mathbf{k}}|^{2}}}-\frac{\nabla^{2}\sqrt{|\Psi_{\mathbf{k}}|^{2}}}{\sqrt{|\Psi_{\mathbf{k}}|^{2}}}\right)|\Phi_{\mathbf{k}}|^{2}d^{3}x.
\end{equation}

\noindent
Since both $\Phi,\Psi$ have the same phase writing $\Phi_{\mathbf{k}}=e^{g(\mathbf{r})}\Psi_{\mathbf{k}}$ where $g(\mathbf{r})$ an unknown real function to be determined yields the integral equation

\begin{equation}
   4\pi \mathbf{Im} f(\mathbf{k},\mathbf{k}')=k\sigma_{sc}-\int \left[\nabla^{2}g(\mathbf{r})+2\nabla g(\mathbf{r})\cdot \nabla lnR+(\nabla g(\mathbf{r}))^{2}\right]e^{2g(\mathbf{r})}R^{2} d^{3}x.
\end{equation}
    
\noindent    
It is clear that we recover the optical theorem for elastic scattering when $g=0$ or $\Phi_{\mathbf{k}}=\Psi_{\mathbf{k}}$ which it is known that expresses the particle conservation requirement but more generally deviations may be observed according to (22) or (23). When the above quantity is positive attenuation of the incident beam is implied. It is interesting to note that the presence of the imaginary potential in (16) introduces non-unitarity since the total particle number is not conserved, but does not affect the reversibility of the scattering amplitude \cite{Glauber} and there is no issue of an effect preceding its cause or retrocausality.

\section{Conclusions}
 
In summary, in the present genuinely multiparticle  formulation ontological priority is given not to the monadic particle but to the emergent several-particle composite. We conclude that the presence of the many particle quantum potential in the modified backward continuity equation suggests non-separability and illustrates the irreducible nature of a multi-particle compound. It is reasonable to deduce then that we cannot examine each particle \textit{singulatim} neither we should consider the quantal multiparticle as a mere mereological aggregate of distinct co-present entities accidentally unified, but rather as a substantial unified whole that constitutes the production of a new physical reality. This particular formalism also ensures that the time anisotropy inferred from the evolution equation and its dependence on terminal values is nomological and intrinsic in nature rather than contingent and accidental.


\begin{thebibliography}{99}
 
 \bibitem{Darwin} Darwin, C. G.: A Collision Problem in the Wave Mechanics.
Proc. R. Soc. Lond. A, \textbf{124}, 375 (1929)

\bibitem{deBroglie} de Broglie, L: The current interpretation of wave mechanics: A critical study. Elsevier Publishing, Amsterdam (1964)
 
\bibitem{EPR} Einstein, A., Podolsky, B., Rosen, N.: Can Quantum-Mechanical Description of Physical Reality Be Considered Complete? 
Phys. Rev., \textbf{47}, 777 (1935)
 
 \bibitem{Bohr} Bohr, N.: Can Quantum-Mechanical Description of Physical Reality Be Considered Complete? Phys. Rev., \textbf{48}, 696 (1935)
 
 \bibitem{Howard} Howard, D.: Einstein on Locality and Separability. Studies in History and Philosophy of Science, \textbf{16}, 171 (1985)
 
\bibitem{Schrodinger} Schr\"{o}dinger, E.: Discussion of Probability Relations between Separated Systems. Mathematical Proceedings of the Cambridge Philosophical Society, \textbf{31}, 555 (1935)
 
\bibitem{Schrodinger2} Schr\"{o}dinger, E.: Probability relations between separated systems. Mathematical Proceedings of the Cambridge Philosophical Society, \textbf{32}, 446 (1936)
 

\bibitem{BohrRosenfeld} Bohr, N., Rosenfeld, L.: On the question of the measurability of electromagnetic field quantities. Reprinted in Wheeler, J.A, Zurek, W.H. (eds.): Quantum theory and measurement, pp. 479-522, Princeton University Press , Princeton, NJ (1983)

\bibitem{Mermin} Mermin, N. D.: What is quantum mechanics trying to tell us?
 Am. J. Phys. \textbf{66}, 753 (1998)
 
 \bibitem{BohmHiley} Bohm, D., Hiley, B.J.: The Undivided Universe: An Ontological Interpretation of Quantum Theory, Routledge (1995)


\bibitem{BohmHiley2}
Bohm, D., Hiley, B. J.: On the Intuitive Understanding of Nonlocality as
Implied by Quantum Theory, Foundations of Physics, \textbf{5}, 93-109, (1975)

\bibitem{Sklar} Sklar, L: Physics and Chance, 
 Cambridge University Press  (1995)
 
 \bibitem{Paternostro} Batalhao, T. B.,  Souza, A. M., Sarthour, R. S., Oliveira, I. S.,  Paternostro, M., Lutz, E.,Serra, R. M.: Irreversibility and the Arrow of Time in a Quenched Quantum System, Phys. Rev. Lett. \textbf{115}, 190601 (2015)
 
 \bibitem{Hawthorne}
Hawthorne, J. and D. Nolan, D: "What Would Teleological Causation Be?", in Hawthorne, J., Metaphysical Essays (Oxford, 2006), pp. 265–83.
 
 \bibitem{Ockendon} Ockendon, J., Howison, S., Lacey, A., Movchan, A.: Applied Partial Differential Equations, Oxford University Press (2003)
 

\bibitem{Gardiner} Gardiner, C. W.:  Stochastic Methods: A Handbook for the Natural and Social Sciences, Springer (2004)

\bibitem{Nuclear} Prinja, A. K,  Larsen, E. W: General Principles of Neutron Transport in  Cacuci, D. G. (Ed.): Handbook of Nuclear Engineering Volume I, Nuclear Engineering Fundamentals, Springer (2010)

\bibitem{Wilmott} Wilmott, P., Howson, S., Howison, S.: The Mathematics of Financial Derivatives: A Student Introduction, Cambridge University Press (1995) 

\bibitem{Renner} Frauchiger, D., Renner, R.: Quantum theory cannot consistently describe the use of itself, Nature Communications, \textbf{9}, 3711 (2018)

\bibitem{Drossel} Drossel, B: arXiv.org quant-ph:1908.10145  (2019)

\bibitem{Gisintime} Gisin, N: "Time Really Passes, Science Can't Deny That", in Time in Physics, eds. R. Renner  S. Stupar, (Springer), pp. 1-15  (2017)

 \bibitem{Feser} Feser, E.: Aristotle's Revenge: The Metaphysical Foundations of Physical and Biological Science, Editiones Scholasticae (2019)

\bibitem{Birula} Bialynicki-Birula, I., Mycielski, J.:  Nonlinear wave mechanics, Annals of Physics \textbf{100}, 62 (1976)

\bibitem{Weinberg2} Weinberg, S.: Testing quantum mechanics, Annals of Physics \textbf{194}, Issue 2, 336 (1989)

\bibitem{Gisin} Gisin, N.: Stochastic Quantum Dynamics and Relativity, Helv. Phys. Acta., \textbf{62}, 363 (1989).

\bibitem{Durt} Colin, S., Durt, T., Willox, W.: de Broglie's double solution program: 90 years later arXiv:1703.06158 (2017)

\bibitem{Smolin} Smolin, L.: Quantum fluctuations and inertia, Physics Letters A,
\textbf{113}, Issue 8, 408 (1986)

\bibitem{Goldin} Doebner, H. D., Goldin, G. A.: On a general nonlinear Schr\"{o}dinger equation admitting diffusion currents, Phys. Lett. A \textbf{162}, 397 (1992).

\bibitem{DG} Tsukagoshi, T., Sugino, O.: Quantum dissipative dynamics using the Doebner-Goldin equation, Physics Letters A \textbf{376}, 3033-3037 (2012) 

\bibitem{Schuch} Schuch, D., Chung, K. M., Hartmann. H.: Nonlinear Schr\"{o}dinger-type field equation for the description of dissipative systems. I. Derivation of the nonlinear field equation and one-dimensional example, Journal of Mathematical Physics \textbf{24}, 1652 (1983)

\bibitem{Kostin} Kostin, M. D: On the Schr\"{o}dinger-Langevin Equation, J. Chem. Phys. \textbf{57}, 3589 (1972)

\bibitem{Wheeler} Wheeler, A.: Law without law reprinted in: Reprinted in Wheeler, J.A, Zurek, W.H. (eds.): Quantum theory and measurement, pp. 182-213, Princeton University Press , Princeton, NJ (1983)

\bibitem{RMP} Sagues. F., Sancho, J. M.,  Garcia-Ojalvo, J.: Spatiotemporal order out of noise, Rev. Mod. Phys. \textbf{79}, 829 (2007)
 
 \bibitem{Hairer} Hairer, M.: A theory of regularity structures. Inventiones Mathematicae \textbf{198}, 269 (2014)
 
 \bibitem{Groessing} Groessing, G: On the Thermodynamic Origin of the Quantum Potential, Physica A: Statistical Mechanics and its Applications, \textbf{388}, 6, 811 (2009)
 
 \bibitem{Durt2} Hatifi, M., Willox, R., Colin, S., Durt, T.: Bouncing Oil Droplets, de Broglie's Quantum Thermostat, and Convergence to Equilibrium, \textbf{20}(10), 780 (2018)
 
 
\bibitem{Scaltsas} Scaltsas, T., Substantial Holism. In: Scaltsas, T., Charles, D., Gill, M.L (eds.): Unity, Identity, and Explanation in Aristotle's Metaphysics, pp. 107-128, Clarendon Press, Oxford (1994); T. Scaltsas, Substances and Universals in Aristotle's Metaphysics Cornell University Press (1994)

 \bibitem{Oderberg} Oderberg, D. S: Real Essentialism, Routledge, New York (2007)
 
\bibitem{Whitehead} Whitehead, A. N: Science and the Modern World, Simon and Schuster (1970)

\bibitem{Bohm} Bohm, D.: Wholeness and the implicate order, Routledge (1980)

\bibitem{Hiley} Hiley, B. J: Some Remarks on the Evolution of Bohm{'}s Proposals for an Alternative to Standard Quantum Mechanics (2010) 
 
 \bibitem{Polyanin} Polyanin A. D., 
Nazaikinskii V. E.: Handbook of Nonlinear Partial Differential Equations, Chapman and Hall/CRC (2016)

\bibitem{Horwich} Horwich, P.: Asymmetries in Time, Cambridge Mass.: MIT Press (1987).
 
\bibitem{Wu} Ohmura, T., Wu, TY.:
Quantum Theory of Scattering
Prentice Hall (1962)

\bibitem{Glauber} R. J. Glauber, High-energy collision theory in "Lectures on Theoretical Physics,"  W. E. Brittin and L. C. Dunham, Eds., Interscience, Vol. 1, New York, 1959, p. 315.



\end{thebibliography}
\end{document}